\documentclass{article}
\usepackage[dvips]{graphicx}
\begin{document}
\title{Self-organization of hierarchical structures in nonlocally coupled replicator models}
\author{Hidetsugu Sakaguchi \\
Department of Applied Science for Electronics and Materials, \\
Interdisciplinary Graduate School of Engineering Sciences,\\
 Kyushu University, Fukuoka 816-8580, Japan
}
\maketitle

\\
\\
Abstract\\
We study a simple replicator model with non-symmetric and nonlocal interactions. Hierarchical structures with prey-predator relations are self-organized from a homogeneous state, induced by the dynamical instability of nonlinear interactions.  
\newpage
Ecological systems and evolution dynamics have been studied as examples of complex systems \cite{rf:1,rf:2}.
The replicator models are often used to study biological evolution among mutually interacting species \cite{rf:3}.
The time evolution of population $x_i$ of the $i$th species obeys an equation 
\begin{equation}
\frac{dx_i}{dt}=x_i(f_{i0}+\sum_jw_{i,j}x_j-c_0),  \,\,{\rm for }\,\,\,i=1,\cdots,N  
\end{equation}
where $N$ is the total number of species, $f_{i0}$ is the natural growth rate of the $i$th species, $w_{i,j}$ denotes the strength of the interaction from the $j$th 
species to the $i$th species, $f_i=f_{i0}+\sum_jw_{i,j}x_j$ is the growth rate for the $i$th species, and $c_0=\sum_{i=1}^Nf_ix_i/\sum_{i=1}^Nx_i$ is the average value of the growth rate. 
The sum of $x_i$ is conserved in the time evolution of Eq.~(1). If the sum is 
assumed to be 1, $x_i$ has a meaning of the population ratio of the $i$th species. The above equation implies that the populations of the species which have the growth rate larger than the average value increase. 
If $w_{i,j}$'s are all zero, only the species with the largest natural growth rate 
survives.  It implies the survival of the fittest species.  
If $w_{i,j}$'s are not all zero, the coexistence of several species is possible.  
Complex dynamics and statistical properties of the replicator models with random interactions have been studied \cite{rf:4}. Steady evolution called "Red Queen" evolution can be modeled with this type of equation \cite{rf:5}. 
We consider the case of $f_{i0}=$const. in this work, that is, all species 
are equivalent by nature in the evolution dynamics.

The population dynamics using Eq.~(1) is closely related with the Lotka-Volterra type equations, which have been more intensively studied in ecological systems \cite{rf:6}.  In the usual Lotka-Volterra models, the interaction strength $\{w_{i,j}\}$'s are usually assumed to be constant. 
They are usually non-symmetric, that is, $w_{i,j}\ne w_{j,i}$.  In the prey-predator relation, where the $j$th species is a prey of the $i$th species and the $i$th species is a predator of the $j$th species,  $w_{i,j}>0$ and $w_{j,i}<0$. In the competitive relation between two species with equivalent strength, the interactions may be symmetric as $w_{i,j}=w_{j,i}<0$. 
However, we can consider some cases where the interaction strengths change 
dynamically. 
For example, there are some models, in which $w_{i,j}$ is a  function of the populations $x_i$ and $x_j$.  
In the population dynamics of trees in forests, competition between trees is not symmetric in general. Larger trees prevent the growth of lower trees, but the lower trees do not seriously affect the larger trees.  Assuming that $x_i$ is the height of a tree at the $i$th position, Yokozawa and Hara proposed a model, in which $W(x_i,x_j)=w_{i,j}x_j$ is expressed as  $W(r)=r^{\beta}$ for $0<r<R_c$ and $W(r)=R_c^{\beta}$ for $r>R_c$, where $r=x_j/x_i$ \cite{rf:7}. 
We proposed a similar form of nonlinear interactions expressed as  
\begin{eqnarray}
w_{i,j}&=&0, \,\,\,\,{\rm for }\,\,r<d_1,\nonumber\\
       &=&a_1(r-d_1), \,\,{\rm  for }\,\,d_1\le r<d_2,\nonumber\\
       &=&a_2(r-d_0), \,\,{\rm  for }\,\,d_2\le r<d_3,\nonumber\\
       &=&a_3(r-d_4), \,\,\,\,{\rm  for }\,\,d_3\le r<d_4,\nonumber\\
       &=&0, \,\,{\rm for }\,\,d_4\ge r,
\end{eqnarray} 
where $r=x_j/x_i$, $d_1=1/d_4,\,d_2=1/d_3,\,d_1<d_2<d_0<1<d_3<d_4$, and $a_1>0,\,a_2<0,\,a_3>0$ \cite{rf:8}.  The interaction $w_{i,j}(r)$ is a continuous and piecewise linear function of the ratio $r=x_j/x_i$. 
This form of interaction seems to be peculiar, but it is one of the simplest 
forms of continuous functions, by which both the mutually equivalent competitive relation and the prey-predator relation can be described.   
The interaction is competitive and equivalent, i.e., $w_{i,j}=w_{j,i}=a_2(1-d_0)<0$, when the population sizes of the $i$th and $j$th species are equal.  Even if the population sizes between the two species are not completely equal but are nearly equal ($d_0<r<(1/d_0)$), the interactions are competitive, i.e., $w_{i,j}<0,\;w_{j,i}<0$.  If the ratio $r=x_j/x_i$ of the population sizes 
is even smaller and satisfies $d_1<r<d_0$, the $j$th species becomes a prey of the $i$th species (that is, the $i$th species is a predator of the $j$th species), i.e.,  $w_{i,j}>0,\,w_{j,i}<0$.  If the ratio $r$ of the population sizes 
is even larger and satisfies $(1/d_0)<r<d_4$, the $i$th species becomes a prey of the $j$th species (that is, the $j$th species is a predator of the $i$th species), i.e.,  $w_{i,j}<0,\,w_{j,i}>0$. If the ratio  $r$ is too large ($r>d_4$) or too small ($r<d_1$), the interaction becomes negligible, i.e.,  $w_{i,j}=w_{j,i}=0$.  The ecological relations change dynamically with the ratio of the population sizes in our model ecological system. 
 Parameter values of 
 $d_0=0.85,\,d_1=3/5,\,d_2=2/3,\,d_3=3/2,\,d_4=5/3,\,a_2=-1.3,\,a_1=a_2(d_2-d_0)/(d_2-d_1)$ and $a_3=a_2(d_0-d_3)/(d_4-d_3)$ are used in the numerical simulations.
A homogeneous solution $x_i=1/N$ for every $i$ is a solution of Eq.~(1), owing to the symmetry of the coupled equations.   
However, the homogeneous solution is not always stable.  The present author, and Yokozawa and Hara found that layered structures are self-organized in a globally coupled model and a locally coupled model with the nearest and next-nearest neighbor interactions, as a result of the instability of the homogeneous solution \cite{rf:7,rf:8}.

In this paper, we study a nonlocally coupled model based on Eq.~(1) with nonlinear interactions of the form Eq.~(2). The position of the $i$th element is located in a one-dimensional lattice or randomly distributed in two dimensions. We assume further that the interaction $w_{i,j}$ takes the value of Eq.~(2) only if the distance between  the $i$th and $j$th elements is smaller than a value determined by a function $f(\bar{x})$, where $\bar{x}$ is the average population size $(x_i+x_j)/2$. We use a simple function  
$f(\bar{x})=\alpha \bar{x}$ in numerical simulations.
The function $f(\bar{x})$ expresses the range of the interaction, and the function $f(\bar{x})=\alpha \bar{x}$ implies that the interaction range increases in proportion to the population size.

Firstly, we study a one-dimensional model. Each species is located at $i$, where $1\le i\le N$. The homogeneous solution is $x_i=x_s=1/N$.
The interaction strength $w_{i,j}$ takes the value of $a_2(1-d_0)$ for a pair of the same population size. 
The linearized equation for a small perturbation $\delta x_i=x_i-x_s$ around the homogeneous solution obeys 
\begin{equation}
\frac{d\delta x_i}{dt}=x_s\{\sum_ja_2(2-d_0)\delta x_j-a_2\sum_j\delta x_i\}, \,\, {\rm for }\,\,i=1,\cdots,N,
\end{equation}
where the sum is taken for the range satisfying $|j-i|<\alpha/N$.
The eigenvalue of the linearized equation can take positive values for $\alpha/N >1$.  For example, if the site number satisfying $|j-i|<\alpha/N$ is 5, 
the eigenvalue $\lambda_k$ for the wavenumber $k$ is expressed as $\lambda_k=x_sa_2\{(2-d_0)(1+2\cos k+2\cos 2k)-5\}$, and therefore $\lambda_k>0$ for $\cos^{-1}\{(-1+\sqrt{5+20/(2-d_0)})/4\}<k<\pi$.
Under the condition, the homogeneous state becomes unstable. 
We have performed numerical simulations for $N=200$.
The initial condition was $x_i=1/N(1+\delta x_i(0))$, where $\delta x_i(0)$ was chosen as a random number between -0.01 and 0.01.
\begin{figure}[htb]
\begin{center}
\includegraphics[width=12cm]{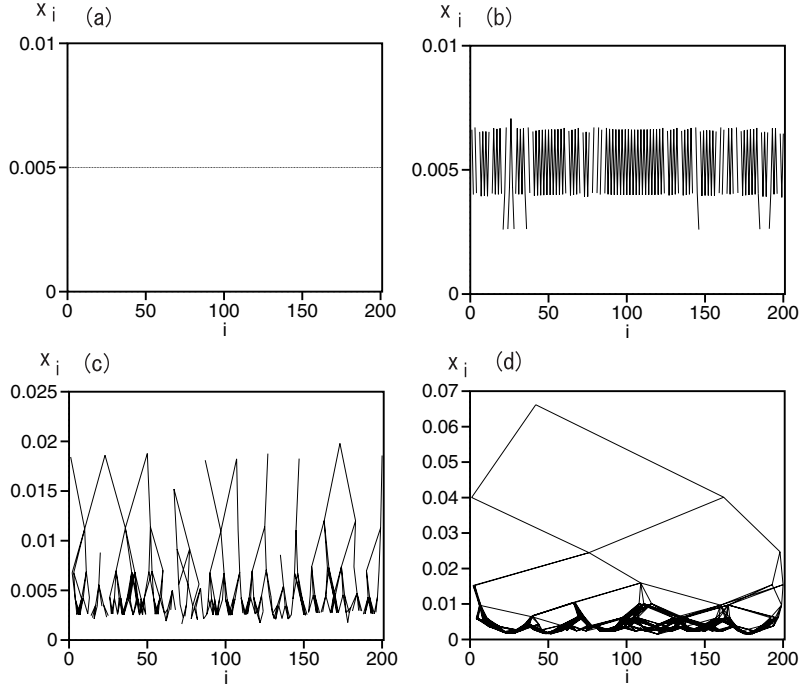}
\caption{Profiles of population sizes $x_i$ for a one-dimensional model Eq.~(1) with $N=200$ and (a) $\alpha=180$, (b) $\alpha=300$, (c) $\alpha=1000$, (d)$\alpha=4000$.
Solid lines denote pairs with  prey-predator relations.}
\label{fig:1} 
\end{center}
\end{figure} 
Figure 1 displays the population size $x_i$ for (a) $\alpha=180$, (b) $\alpha=300$, (c) $\alpha=1000$, (d) $\alpha=4000$ at $t=20000$.
Solid lines denote pairs with prey-predator relations, in which $w_{i,j}<0$ and $w_{j,i}>0$ are satisfied. 
The homogeneous state is stable for $\alpha=180$, since there are no interactions except for the self interaction expressed by $w_{i,i}$.
The homogeneous state becomes unstable for $\alpha/N>1$ and zigzag structures appear for $\alpha=300$. As $\alpha$ is increased, hierarchical structures with higher levels are created. 
Strong elements appear and dominate locally the regions of their population sizes.  The local winners have larger population sizes and therefore larger interaction ranges.  Competitive relations are therefore created among the local winners, and the higher hierarchical relations are created by the competition among the local winners like a tournament.  For $\alpha=4000$, there appears an element which dominates the whole region, and a hierarchical structure from top to bottom is created. 
Which element dominates the whole region depends on the initial conditions.

Next, we consider a two-dimensional model.
The position of the $i$th element is randomly distributed inside 
a circle of radius 10.
\begin{figure}[htb]
\begin{center}
\includegraphics[width=15cm]{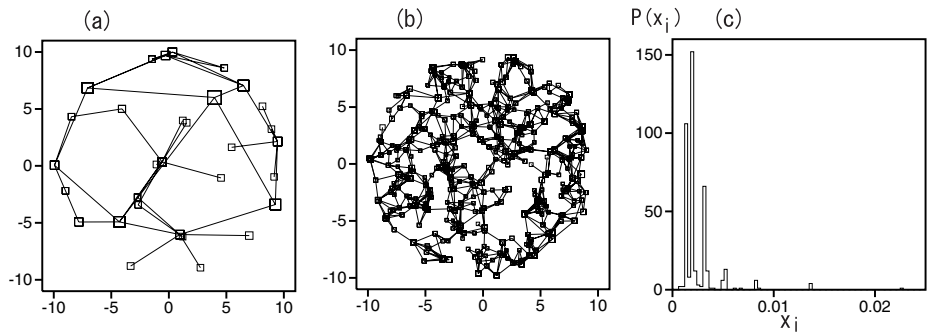}
\caption{Hierarchical structure for Eq.~(1) with $N=400$ in two dimensions. Each element is randomly distributed in a circle of radius 10. The parameter value of $\alpha=800$. Figure 2(a) is the network with larger size populations, which satisfies $x_i>0.005$, and (b) is the network with the smaller size populations. The area of the square at each site is proportional to $x_i$.
Figure 2(c) is a histogram of the population size.}
\label{fig:2} 
\end{center}
\end{figure} 
The population size at the $i$th element is denoted as $x_{i}$.
The system size is $N=400$.
The initial conditions are $x_{i}=1/N(1+\delta x_{i}(0))$, where $\delta x_{i}(0)$ is a random number between -0.01 and 0.01.
As a result of the time evolution by Eq.~(1), a two-dimensional hierarchical structure is naturally created. 
As the parameter value $\alpha$ is increased, we have observed development of  hierarchical structures as in the one-dimensional model. 
Figure 2 displays a developed hierarchical network for $\alpha$ is 800. 
The prey-predator relations are displayed in this figure.
To show the network clearly, we have classified the set of $i$ into two groups.  Figure 2(a) is a network for larger size populations, which satisfies $x_i>0.005$, and Fig.~2(b) is a network for the smaller size populations, which satisfies $x_i<0.005$.
The population size $x_i$ is expressed by the area of the square located 
at the $i$th position. 
The elements, which have larger populations, dominate the larger regions.
The average number of preys for each element is 4.3 in this simulation.
Figure 2(c) displays a histogram of population size.  
It is seen from this histogram that a layered structure with 7 levels is created.

To summarize, we have proposed a nonlocally coupled replicator model for the self-organization of hierarchical structures.
The interaction range was assumed to be proportional to the population size, and the nonlinear interactions between two species are assumed to depend on the ratio of the population sizes. 
Owing to the two assumptions, a hierarchical structure of the interaction network is self-organized. 
The parameter $\alpha$ determines the level number of hierarchical structures.
For large $\alpha$, a self-similar-like hierarchical structure is created, in which the population sizes and interaction ranges increase geometrically as the level is increased.  
The stationary solutions to Eq.~(1) are not unique as shown in a globally coupled model \cite{rf:8}, however, similar forms of hierarchical structures were 
obtained from different initial conditions.
 If the initial condition is completely localized as $x_{i}=1$ for $i=N/2$ and $x_i=0$ for the other $i$, the localized state is maintained in the time evolution of Eq.~(1), since new populations cannot be born from zero. 
As a natural modified model,  a  diffusion process may be introduced in the population dynamics as
\begin{equation}
\frac{dx_i}{dt}=x_i(\sum_jw_{i,j}x_j-c_0)+D(x_{i+1}-2x_i+x_{i-1}),  \,\,{\rm for }\,\,i=1,\cdots,N,  
\end{equation}
where $D$ is a diffusion constant. The diffusion process corresponds to the migration  in ecosystems.
New populations are migrated at the sites where $x_i=0$ originally, and then the localized state becomes unstable. We have checked that similar hierarchical structures are naturally created  from the localized initial conditions even for very small $D$. 
We have used a special form of interactions. However, the dynamical instability 
of the homogeneous state is due to the negative derivative of $dw_{i,j}(r)/dr$ at $r=1$ and 
other forms of interactions with the same characteristics can induce similar 
hierarchical structures.   
The self-organization of the hierarchical structure owing to mutual interactions is suggestive of the formation of complex networks in ecosystems, economic and social systems \cite{rf:9}.

\end{document}